\documentclass[aps,10pt,pra,twocolumn,groupedaddress,floatfix,superscriptaddress,showpacs,showkeys,amsfonts]{revtex4-1}
\usepackage{amsfonts}
\usepackage{amsmath}
\usepackage{amssymb}
\usepackage{graphics,graphicx}
\graphicspath{{./}}
\usepackage{epsf}
\usepackage{subfigure}
\usepackage[%
  colorlinks=true,
  urlcolor=blue,
  linkcolor=blue,
  citecolor=blue
]{hyperref}
\usepackage[all]{hypcap}
\usepackage{color}
\usepackage[utf8]{inputenc}
\newcommand{\be}{\begin{equation}}
\newcommand{\ee}{\end{equation}}
\newcommand{\bea}{\begin{eqnarray}}
\newcommand{\eea}{\end{eqnarray}}
\newcommand{\eq}[1]{Eq.~\eqref{#1}}

\newcommand{\eqss}[2]{Eqs.~\eqref{#1}-\eqref{#2}}

\newcommand{\fig}[1]{Fig.~\ref{#1}}

\newcommand{\bem}{\begin{multline}}
\newcommand{\eem}{\end{multline}}

\newcommand{\apporsm}[1]{App.~\ref{#1}}
\newcommand{\appredef}[2]{#2}

\graphicspath{ {figures/} }

\AtBeginDocument{%
    \newwrite\bibnotes
    \def\bibnotesext{Notes.bib}
    \immediate\openout\bibnotes=\jobname\bibnotesext
    \immediate\write\bibnotes{@CONTROL{REVTEX41Control}}
    \immediate\write\bibnotes{@CONTROL{%
    apsrev41Control,author="08",editor="1",pages="0",title="0",year="1"}}
     \if@filesw
     \immediate\write\@auxout{\string\citation{apsrev41Control}}%
    \fi
}
\begin{document}

\title{Multistability of Driven-Dissipative Quantum Spins}

\author{Haggai Landa}\email{haggaila@gmail.com}
\affiliation{Institut de Physique Th\'{e}orique, Universit\'{e} Paris-Saclay, CEA, CNRS, 91191 Gif-sur-Yvette, France}
\author{M. Schir\'o}
\email{marco.schiro@ipht.fr}\thanks{On Leave from: Institut de Physique Th\'{e}orique, Universit\'{e} Paris Saclay, CNRS, CEA, F-91191 Gif-sur-Yvette, France}
\affiliation{JEIP, USR 3573 CNRS, Coll\'ege de France,   PSL  Research  University, 11,  place  Marcelin  Berthelot,75231 Paris Cedex 05, France}
\author{Gr\'egoire Misguich}\email{gregoire.misguich@cea.fr}
\affiliation{Institut de Physique Th\'{e}orique, Universit\'{e} Paris-Saclay, CEA, CNRS, 91191 Gif-sur-Yvette, France}
\affiliation{Laboratoire de Physique Th\'{e}orique et Mod\'{e}lisation, CNRS UMR 8089, Universit\'{e} de Cergy-Pontoise, 95302 Cergy-Pontoise, France}

\begin{abstract}
We study the dynamics of lattice models of quantum spins one-half, driven by a coherent drive and subject to dissipation.
Generically the meanfield limit of these models manifests multistable parameter regions of coexisting steady states with different magnetizations. We introduce an efficient scheme accounting for the corrections to meanfield by correlations at leading order, and benchmark this scheme using high-precision numerics based on matrix-product-operators in one- and two-dimensional lattices. Correlations are shown to wash the meanfield bistability in dimension one, leading to a unique steady state. In dimension two and higher, we find that multistability is again possible, provided the thermodynamic limit of an infinitely large lattice is taken \emph{first} with respect to the long time limit. Variation of the system parameters results in jumps between the different steady states, each showing a critical slowing down in the convergence of perturbations towards the steady state. Experiments with trapped ions can realize the model and possibly answer open questions in the nonequilibrium many-body dynamics of these quantum systems, 
beyond 
the system sizes accessible to present numerics.
\end{abstract}

\maketitle

Coherent control over quantum single- and few-body dynamics is continuously improving, spanning atomic, optical, and solid-state systems \cite{haroche_exploring_2006, yamamoto_semiconductor_2000, schoelkopf_wiring_2008}. 
An ongoing effort is focused on assembling many  individually tunable systems and studying  the ensuing many-body dynamics.
A significant challenge lies in realizing unitary dynamics, however, the inevitable presence of dissipative processes can
be utilized in different scenarios, such as by reservoir engineering \cite{bardyn2013topology}. Coherent time-periodic driving is a useful tool \cite{annurevOkaKitamura}, and rich dynamics are observed with systems of strong light-matter interactions at the interface between quantum optics and condensed matter  \cite{Greentree_NatPhys2006,Hartmann_NatPhys2006,ChangEtAlNatPhys06,Koch_PRA11, Petrescu_PRA12, Angelakis_FQH_PRL08,umucalilar_artificial_2011, HafeziLukinTaylor_NJP2013, rechtsman2013photonic, 
OtterbachEtAl_PRL13,CarusottoCiutiRMP13,LeHurReview16,
NohAngelakisRepProgPhys2016,HartmannJOpt2016,SchiroPRL16,FitzpatrickEtAlPRX17,FinkEtAlPRX17}. 
Systems with a competition between interactions, nonlinearity, coherent external driving and dissipative dynamics include arrays of coupled circuit quantum electrodynamic units~\cite{AndrewNatPhys,SchmidtKochAnnPhy13}, 
cold atoms~\cite{BlochDalibardNascimbeneNatPhys12}, and ions \cite{bohnet2016quantum}. 
Critical phenomena and dissipative phase transitions in these open systems often come with new properties and novel dynamic universality classes~\cite{Tomadin_prl10,SiebereHuberAltmanDiehlPRL13,LeeEtAlPRL13,JinetalPRL13,MarinoDiehlPRL16,
ScarlatellaFazioSchiroPRB19,munoz2019spontaneous}.

The state of an open quantum system is defined by a density matrix $\rho$, with the dynamics often treated using a Lindblad master equation, describing a memory-less bath, and the time evolution generated by the Liouvillian superoperator acting on $\rho$~\cite{BreuerPetruccione}.
The theoretical tools available for
open quantum many-body systems are relatively limited. For driven-dissipative lattice models the meanfield (MF) approach is often employed,
with $\rho$ approximated as a product of single-site density matrices. The dynamics of local observables are described by nonlinear equations, studied, e.g., for lattice Rydberg atoms \cite{lee_antiferromagnetic_2011,qian_phase_2012, 
marcuzzi2014universal, CarrEtAlPRL13,LetscherEtAlPRX17,ParmeeEtAlPRA18}, coupled quantum-electrodynamics cavities and circuits \cite{PhysRevLett.110.163605,jin_steady-state_2014,SchiroPRL16}, nonlinear photonic models \cite{foss-feig_emergent_2017,biondi_nonequilibrium_2017}, and spin lattices \cite{chan_limit-cycle_2015,wilson_collective_2016}.
A key feature of the MF phase diagrams are multistable parameter regions where two or more
steady states coexist.

However, the Lindblad equation converges in general to a unique steady state in finite systems \cite{spohn1977algebraic,PhysRevA.98.042118}, making the status of the MF approximation unclear. Indeed, significant deviations from MF have been found using approximation schemes accounting for quantum correlations \cite{weimer_variational_2015,biondi2017spatial,JinEtAlPRX16},
 and also using exact numerical methods (quantum trajectories~\cite{daley2014quantum} and Matrix Product Operators (MPO)~\cite{prosen_matrix_2009}). 
In one-dimensional (1D) lattices with nearest-neighbour (NN) interactions, the MF bistability is found to be replaced by a crossover driven by large quantum fluctuations \cite{weimer_variational_2015,PhysRevA.93.023821,foss-feig_emergent_2017,vicentini_critical_2018}. 
In contrast, in certain 2D NN models, MF bistability has been found by approximate methods to be replaced by a first-order phase transition between two states, for nonlinear bosons using a truncated Wigner approximation \cite{foss-feig_emergent_2017,vicentini_critical_2018}, and for Ising spins using a variational ansatz acounting for short-range correlations \cite{weimer_variational_2015}, a cluster MF approach~\cite{JinEtAlPRB18} and two-dimensional tensor network states~\cite{KshetrimayumEtalNatComm2017}. In a parameter region around the jump, the convergence towards the steady state slows down \cite{weimer_variational_2015,vicentini_critical_2018},
a phenomenon related to a gap closing in the spectrum of the Liouvillian \cite{cai2013algebraic,macieszczak_towards_2016, PhysRevA.98.042118}.

In this Letter we study a driven-dissipative model of spins one-half with XY (flip-flop) interactions  in presence of coherent drive and dissipation, using a combination of MPO simulations and an approximation scheme which accounts for quantum fluctuations beyond meanfield (MFQF). For one dimensional lattices we confirm the existence of a unique steady state in the thermodynamic limit.  As our main result, we find that in dimension two and higher multistability (in particular, bistability) is again possible, with jumps between the different steady states accompanied by a critical slowing down, provided that the thermodynamic limit of an infinitely large lattice is taken \emph{first} with respect to the long time limit. We argue that this order of limits is physically plausible, and we link the bistability to the fact that for finite size and time the probability distribution of relevant observables develops a strong 
bimodal structure. Depending on the order of limits, bimodality leads either to a first-order dissipative phase transition (as usually discussed when the long time limit is taken first), or to a bistable regime. We thus provide a theoretical scenario reconciling our results with the literature cited above, and finding similar dynamics in a model with Ising interactions (see below and \apporsm{App:Ising}) indicates the generality of our results. We suggest that in an experimental platform based on trapped-ion quantum simulators, such a question can be addressed.

{\it Model.}
 We consider a  quantum system with $N$ sites $R\in \mathbb{Z}^D$ on a hypercubic lattice in $D$ spatial dimensions, for which the connectivity is $\mathcal{Z}=2D$. The master equation for $\rho$ is defined using the Liouvillian $\hat{\mathcal{L}}$,
 \be \partial_t\rho = \hat{\mathcal{L}}[\rho]\equiv -i[H,\rho]+\mathcal{D}[\rho],\qquad \hbar=1.\label{Eq:dtrho}\ee
The Hamiltonian describing Rabi oscillations of two-level systems with a drive detuned by $\Delta$ from the resonant transition frequency and a Rabi frequency $\Omega$, is given in a frame rotating with the drive by 
 \be H=\sum_{R}\left[\frac{\Delta}{2}\sigma_R^z+\Omega\sigma_R^x\right] -\sum_{\langle R,R'\rangle} J\left(\sigma^+_R \sigma^-_{R'} +{\rm h.c.}\right) ,\label{Eq:HR}\ee
where the second sum extends over all pairs of NN sites, describing hopping with amplitude $J$, with spin-$\frac{1}{2}$ operators (Pauli matrices) $\sigma_R^a$, $a=\{x,y,z\}$, and
%ladder operators
$\sigma^{\pm}_R = ({\sigma^{x}_R\pm i\sigma^y_R})/{2}$.
For spin losses occurring independently at each site with rate $\Gamma=1$ (which fixes the frequency and time units),
\be
 \mathcal{D}[\rho]=\sum_R\left[ \sigma_R^-\rho \sigma_R^{+}-\frac{1}{2}\left(\sigma_R^{+}\sigma_R^- \rho + \rho\sigma_R^{+}\sigma_R^- \right)\right].\label{Eq:Dissipator}\ee
 Aside from translation invariance, this model has no manifest microscopic symmetries. Its MF phase diagram displays bistability \cite{PhysRevA.93.023821,wilson_collective_2016} in a region of parameters that terminates at a second order point where an emerging $Z_2$ symmetry spontaneously breaks~\cite{marcuzzi2014universal}. In 1D the steady state is unique as obtained by MPO simulations \cite{PhysRevA.93.023821}. Here we focus on higher dimensions, which we find to manifest bistability in the thermodynamic limit.
%\new{The important observation that we must make now is that the order of taking the two limits $t\to\infty$ and $N\to\infty$, is important in understanding the dynamics.}

{\it Dynamics of Observables.}
From the master equation one can derive a hierarchy of equations of motion for $n$-points expectation values
%of operators
of the form $\langle \sigma_{R_1}^{a}\sigma_{R_2}^{b}\cdots \sigma_{R_n}^{c}\rangle\equiv {\rm tr}\{\rho\sigma_{R_1}^{a}\sigma_{R_2}^{b}\cdots \sigma_{R_n}^{c}\}$, which depend on the value of correlators at the next order, $n+1$. Assuming a translationally-invariant  density matrix, we define the uniform vector mean
magnetization,
$ \mu_a(t)= \left\langle \frac{1}{N}\sum_R\sigma_R^a\right\rangle= \left\langle\sigma_R^a\right\rangle$,
and its equations of motion 
\bea \partial_t\mu_x &= &-{J}\mathcal{Z} [\mu_{y}\mu_z+\eta_{yz}(1)] -{\Delta} \mu_y - \mu_x/2,\quad\label{Eq:eomRmag_x} \\
\partial_t\mu_y &= &{J}\mathcal{Z} [\mu_{x}\mu_z+ \eta_{xz}(1)] -2\Omega\mu_z +{\Delta} \mu_x - \mu_y / 2,\quad \label{Eq:eomRmag1} \\ 
\partial_t \mu_z &= &2 \Omega \mu_y - \left(1 +\mu_z\right)\label{Eq:eomRmag2}.\quad
\eea
with the connected two-point correlation functions,
\bem {\eta}_{ab}(R,R',t)\equiv \left\langle \left(\sigma_{R}^a -\mu_a\right)\left(\sigma_{R'}^b-\mu_b\right)\right\rangle\\=\left\langle \sigma_{R}^a\sigma_{R'}^b\right\rangle - \mu_a\mu_b,\quad R\neq R',\label{Eq:etadef0}\end{multline}
and setting $R'=0$ using the translation invariance, ${\eta}_{ab}(1)$ is the correlator at a NN of the origin.

Equations \eqref{Eq:eomRmag_x}-\eqref{Eq:eomRmag2} are exact. The limit ${\eta}\to 0$ reduces $\rho$ to a product of identical on-site states, leading to the MF equations,
whose steady state and dynamics are studied in detail in \cite{etatheory}.
We present an approximate scheme going beyond MF, formally based on an expansion in $1/\mathcal{Z}$ (with a related approach in \cite{biondi2017spatial}).
Neglecting the connected three-point correlators
$\left\langle \left(\sigma_{R}^a -\mu_a\right)\left(\sigma_{R'}^b-\mu_b\right)\left(\sigma_{R''}^c-\mu_c\right)\right\rangle\approx 0$,
% $R\neq R'\neq R''$
allows us to derive (see \apporsm{App:eta}) coupled equations for $\eta_{ab}(R,t)$, which we solve numerically together with their feedback into \eqss{Eq:eomRmag_x}{Eq:eomRmag2}.
Since the short-range correlators $\eta_{ab}(1)$ appearing in \eqss{Eq:eomRmag_x}{Eq:eomRmag2} are dynamically coupled to all distances in the lattice, the MFQF method accounts for the spatial structure of correlation functions. The simulations have been verified to converge as a function of $N$,
%\new{
and hence we can approximate the system dynamics as a function of time with the limit $N\to\infty$ taken first.
%}

\begin{figure}
\includegraphics[width=3.2in]{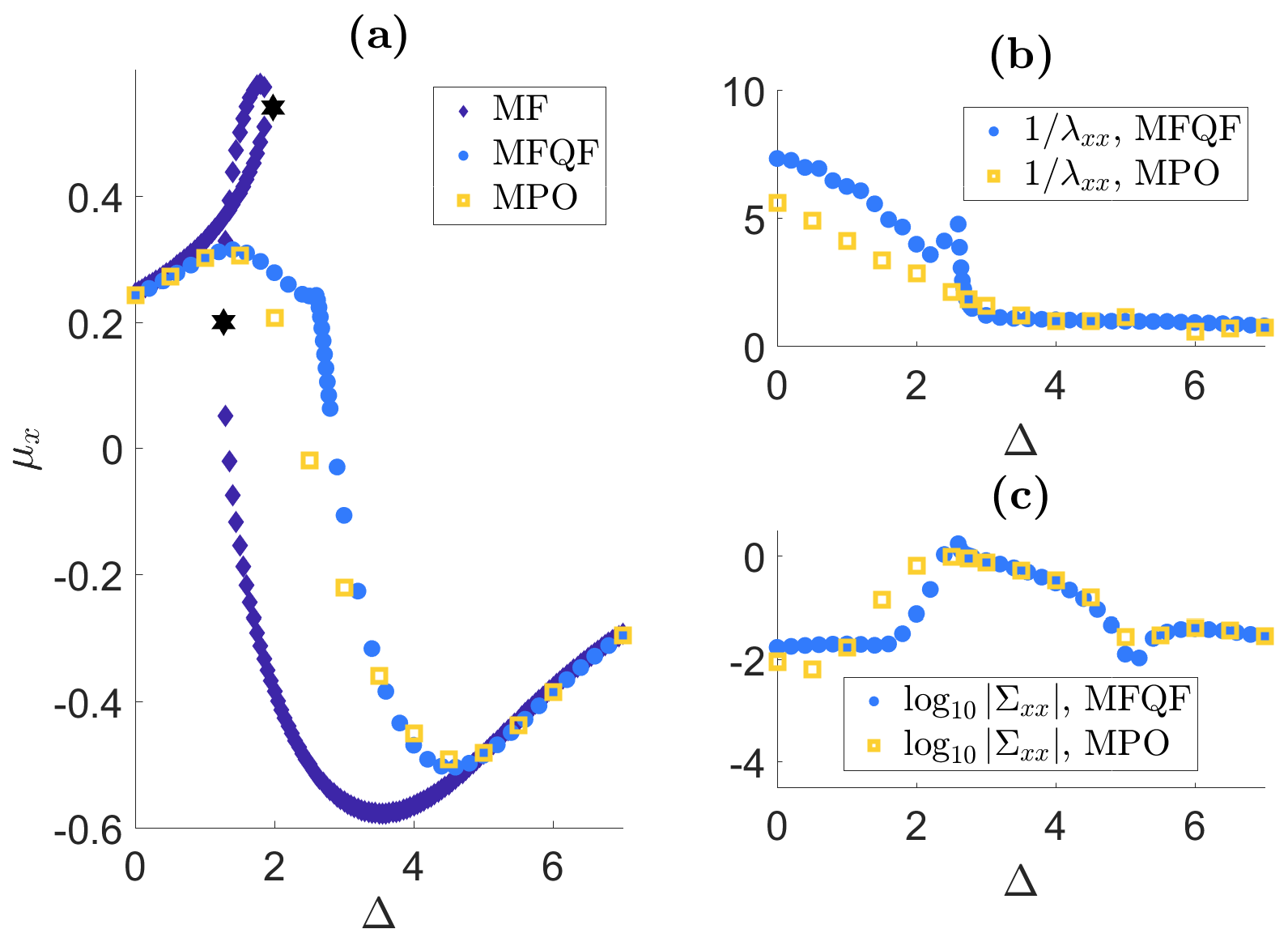}
\caption{(a) Mean steady-state $x$ magnetization $ \mu_x^S$ as a function of $\Delta$ for $\Omega=0.5$ and $J\mathcal{Z}=4$, on a 1D lattice. The meanfield (MF) limit manifests bistability, with three co-existing solutions, two of which -- those on the branches coming from the limits of $\Delta\to\{0,\infty\}$, are stable. Two black hexagrams mark the points where the unstable branch meets each of the two stable ones. An exact numerical treatment using Matrix Product Operators (MPO) shows a crossover within a range of $\Delta$ shifted from the MF bistability region. An approximation incorporating quantum fluctuations at leading order (MFQF) follows approximately the MPO result in a large range of parameters. (b) The correlation length $\lambda_{xx}$ defined by fitting $\eta_{xx}\sim \exp\{-\lambda_{xx}R\}$, and (c) the total correlation $\Sigma_{xx}= \sum_R\eta_{xx}(R)$, calculated in MPO and MFQF, showing that the latter approximation is capable of capturing the spatial structure and relative magnitudes of 
two-point correlations in the lattice.
} \label{Fig:Comparison}
\end{figure}

%\new{
{\it Correlations wash away Bistability in 1D.} We start our analysis with numerically exact MPO calculations of large lattices in 1D. The density matrix $\rho$ can be considered as a pure state in an enlarged
Hilbert space with four states per site \cite{mascarenhas_matrix-product-operator_2015}, allowing us to solve the Lindblad evolution using a method formally similar to pure state unitary evolution encoded using well-established matrix product states (see \cite{SchollwockAnnPhys11,zaletel_time-evolving_2015} and references therein).
We evolve $\rho(t)$ in a 1D chain with open BC (translation invariance is not enforced), using an MPO algorithm \cite{prosen_matrix_2009,benenti_charge_2009, mascarenhas_matrix-product-operator_2015}, with an implementation based on the iTensor library \cite{itensor}, a Trotter decomposition of order four \cite{zaletel_time-evolving_2015,bidzhiev_out--equilibrium_2017}, and bond dimension $\chi=300$.
With up to 200 spins we checked that observables measured in the central region of the chain had negligible finite-size effects and truncation errors at the scale of the plots, allowing us to obtain their steady-state bulk values corresponding to the thermodynamic limit.
%}

 Figure \ref{Fig:Comparison}(a) shows the $x$ component of the steady-state magnetization, $\vec\mu^S\equiv \lim_{t\to\infty}\vec\mu(t)$, in 1D, for $J\mathcal{Z}=4$ as a function of $\Delta$. In MF, $\vec\mu^S$ is unique except for $1.3 \lesssim \Delta\lesssim 1.9$, 
where there are two co-existing stable solutions in addition to an unstable solution.
At the presence of quantum correlations (in MPO and MFQF), the magnetization departs significantly from the MF prediction, with a crossover between the two limiting regimes, in the range $1.5\lesssim\Delta \lesssim 5$.
We define the inverse correlation lengths $\lambda_{ab}$ by fitting the six correlation functions to $\eta_{ab}(R) \sim \exp\{-\lambda_{ab}R\}$.
For simplicity, we present in \fig{Fig:Comparison}(b) only one correlation length, and \fig{Fig:Comparison}(c) shows the corresponding total correlation measured by $\Sigma_{ab}= \sum_R\eta_{ab}(R)$. The spatial structure of the two-point correlation
undergoes a sharp change within the crossover region, from relatively small but widely extended correlations for low $\Delta$, to much larger but very short-ranged correlations, for high $\Delta$. A separate analysis of the correlations shows that at the same time, the correlations change nature from periodic modulations (spin density-wave character), to being overdamped in space.

\begin{figure}
\includegraphics[width=3.2in]{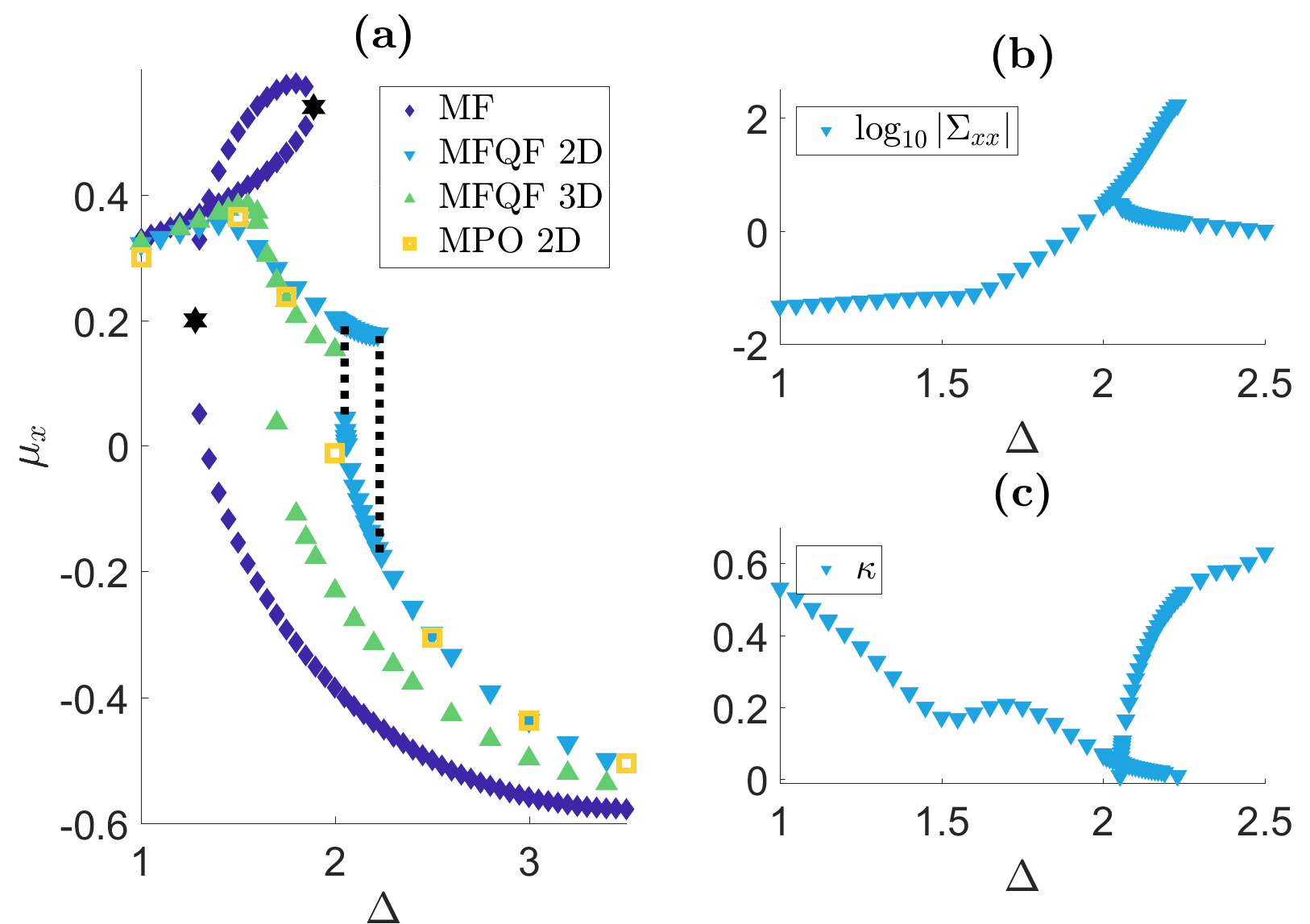}
\caption{(a) Mean steady-state $x$ magnetization $\mu_x^S$ in the MFQF $N\to\infty$ approximation in 2D-3D, together with the MF limit and
2D-MPO results (for a $12\times4$ cylinder). The parameters are as in \fig{Fig:Comparison}, with ${J}\mathcal{Z}$ kept fixed by varying $J$ with the dimension. For $D\ge 2$ MFQF predicts multistability, with two stable branches approaching the MF branches in an increasingly larger parameter region as $D$ is increased. The dotted black lines indicate the edges of the 2D bistable region. The simulations were run with lattices of up to $200^2$ and $40^3$  sites and periodic BC. (b) The total correlation in 2D, showing a difference of up to two orders of magnitude in the bistable phases. (c) The rate of convergence to the 2D steady states, fitted to an exponential form $\sim e^{-\kappa t}$, showing a critical slowing down of the dynamics as $\kappa \to 0 $ at the two bistability edges.} \label{Fig:Comparison2}
\end{figure}

As \fig{Fig:Comparison} shows, the MFQF approximation correctly captures the uniqueness of the steady state and the disappearance of bistability in 1D.
On both sides of the crossover region, the results are quantitatively accurate.
In its center, the approximation reaches too large values for $\eta_{ab}(R)$ and the correlation length. More generally, we find that as $J$ is increased in 1D, the MFQF approach loses its accuracy (for parameters of strong correlations), plausibly because of the role of higher-order correlation functions that are neglected, which can lead at much larger $J$ to the breakdown of the approximation. However,
the MFQF approach is
easy to generalize to higher dimensions, and
quantitatively accurate in regions with moderate correlations.

{\it Bistability in Higher dimensions.} Figure \ref{Fig:Comparison2}(a) shows the results of simulations with large 2D and 3D lattices, for $J\mathcal{Z}=4$.
%\new{
The MFQF theory, that allows simulating large lattices (with $N\to\infty$), is compared with 2D-MPO calculations, limited to a finite-size system, for which,
%}
as in 1D, $\rho$ is encoded as
a product of matrices. The matrix product runs over
a snake-like path visiting all the sites of a cylinder of length $L_x=12$ and perimeter $L_y=4$ (see \apporsm{App:2D-MPO}).
Such an approach has been applied in ground-state calculations
of 2D models \cite{stoudenmire_studying_2012}, but we are not aware of previous 2D-MPO Lindblad calculations.
For $\Delta\lesssim1.5$ and $\Delta\gtrsim2.5$ the agreement between MPO and MFQF is almost perfect,
giving a nontrivial check of the ability of MFQF to capture significant correlation effects (that result in $\vec \mu$ strongly departing from MF). 
The computational cost of guaranteeing a high accuracy in 2D-MPO calculation is exponential in $L_y$ (see \apporsm{App:2D-MPO}), limiting the present MPO calculations to relatively small systems, which cannot show bistability (and a possible discontinuity would also be smeared out).

As our main result, using MFQF we find in 2D two stable $\vec\mu^S$ branches, that in 3D extend over larger ranges of $\Delta$, converging towards the MF bistability region and magnetization values. Figure \ref{Fig:Comparison2}(b) shows that $\Sigma_{xx}$ increases by two orders of magnitude for one of the bistable states, and 
 \fig{Fig:Comparison2}(c) shows the asymptotic relaxation rate
associated to the convergence to $\vec\mu^S$. It is obtained by fitting
%the derivative of the mean magnetization magnitude
$\partial_t \vec\mu^2\sim e^{-\kappa t}$ at large times ($t\sim 100$).
The fact that $\kappa\to 0$ at the branch edges in 2D indicates a critical slowing down when approaching the end of the bistability region in the phase that is about to disappear, leading to a discontinuous jump.
The MFQF approach does however not {\em always} predict bistability in 2D.
Replacing each hopping term in \eq{Eq:HR} by the Ising coupling $J_z\sigma_R^z\sigma_{R'}^z$, we find a smooth crossover for moderate $J_z$ (as obtained using cluster meanfield \cite{JinEtAlPRB18}),
and a small bistability region for stronger couplings (again as in \cite{JinEtAlPRB18});
See \apporsm{App:Ising}.

{\it Bistability, Liouvillian spectrum and bimodality.}
We henceforth return to the question raised in the introduction:  how to reconcile the uniqueness of the steady state in finite systems, with bistability seen when taking first the thermodynamic limit of infinite size, and then the long-time limit?

Considering the Liouvillian $\hat{\mathcal{L}}$ of \eq{Eq:dtrho},
the unique thermodynamic steady state corresponds to $\rho_{ss}=\lim_{N\to \infty} \lim_{t\to\infty} \rho(t)$, which is independent of the initial conditions, and is an eigenstate of $\hat{\mathcal{L}}$ at any $N$, $\hat{\mathcal{L}}\rho_{ss}=0$.
Assuming bistability, 
we define $\rho_1$ and $\rho_2$ as the two distinct density matrices obtained from $\lim_{t\to \infty} \lim_{N\to\infty} \rho(t)$, which depend on the initial conditions. For $N$ large but finite the bistability should be replaced by long-lived metastable states, in which case $\rho_1$ and $\rho_2$ are defined at times $t\gg 1/\Gamma$, but small compared to the lifetime of these metastable states.
As in the model studied here, these two states have different local properties: a local
observable ({\it i.e.} sum of local terms, e.g.~$M_x=\frac{1}{N}\sum_R \sigma^x_R $)  has a probability distribution $P_1(m)$
with a single peak centred around $m_1$ in $\rho_1$, and a distribution $P_2(m)$ peaked around $m_2$($\ne m_1$) in $\rho_2$.
At the same time, metastability implies some relaxation time diverging with $N$, and the spectrum of $\hat{\mathcal{L}}$ must have at least one nonzero eigenvalue $\lambda_E$ with a vanishingly small real part, $\lim_{N\to\infty}{\rm Re}(\lambda_E)=0$, other eigenvalues being separated by a gap $\mathcal{O}(\Gamma)$ \cite{macieszczak_towards_2016,rose_metastability_2016}.
We assume for simplicity that $\hat{\mathcal{L}}$ has a unique such small eigenvalue (therefore real), and denote by $\rho_E$ the associated eigenstate (or  eigenmatrix).
$\rho_{ss}$ and $\rho_E$  are the eigenstates from which all long-lived states can be  constructed, since for $t$ much larger than $1/\Gamma$ we can ignore higher ``excited'' eigenstates. So, for $\rho_1$ and $\rho_2$ to be long-lived, they must be linear combinations of  $\rho_{ss}$ and $\rho_E$. As physical states have a trace equal to 1, and since ${\rm Tr}\rho_{ss}=1$ and ${\rm Tr}\rho_{E}=0$, there must exist two distinct scalars $a_1$ and $a_2$ such that $\rho_i=\rho_{ss}+a_i\rho_E$ with $i=1,2$ \cite{macieszczak_towards_2016,rose_metastability_2016, PhysRevA.98.042118,LetscherEtAlPRX17}.
Inverting these relations we get $\rho_{ss}=(a_2\rho_1-a_1\rho_2)/(a_2-a_1)$.
So, if $a_1$ and $a_2$ are both nonzero (which may not be always the case) $\rho_{ss}$ is a ``cat state'' (with correlation functions extending over the system size), being a linear combination of two uniform physical states with different local properties. 
%Note that, with the additional assumption that $\rho_{i=1,2}$ are associated to some spontaneous breakdown of a $\mathbb{Z}_2$ symmetry, it is possible to be more precise and establish \cite{PhysRevA.98.042118} that $\rho_{ss}=\frac{1}{2}\left(\rho_1+\rho_2\right)$.
In $\rho_{ss}$, the probability distribution $P_{\rm ss}=(a_2 P_1 - a_1 P_2)/(a_2-a_1)$ of a local observable is {\em bimodal}, peaked around the two mean values $m_{1,2}$ realized in the states $\rho_{i=1,2}$.

Using exact diagonalization on small systems we have computed such distributions for the fully-connected (FC) version of the present XY model, which is bistable in the thermodynamic limit
(where MF becomes exact), and for the 1D and 2D cases. We find (\apporsm{App:Bimodality}),
that the magnetization becomes bimodal in parts of the MF bistability region for the FC and 2D cases, whereas it stays mono-modal in 1D. The scaling with $N$ of the bimodal peaks is beyond the scope of the current work, however, the mean value of an observable computed with $\rho_{ss}$ may become {\em discontinuous} as $N\to\infty$ at some value of the parameter \cite{foss-feig_emergent_2017,casteels_critical_2017, vicentini_critical_2018,PhysRevA.98.042118}.
This could correspond, in the discussion above, to smoothly varying  $\rho_{i=1,2}$ but discontinuous jumps of $a_1$ and $a_2$.
Hence, a unique steady state with a discontinuous jump
is a-priori compatible with bistability and hysteresis and, in the present scenario, finding one or the other
in a theory calculation is a matter of order of limits.

Moreover, since the support of $P_1$ has essentially no overlap with that of $P_2$ (for a large enough system), any density matrix which is not a convex combination of $\rho_1$ and $\rho_2$ would give some (unphysical) negative probability density. This means that all physical long-lived states are convex combinations of the mono-modal states $\rho_i$, and
the latter thus coincide with the extreme states of \cite{macieszczak_towards_2016}.
The above discussion therefore connects our results both with the theories of first-order phase transitions, and the theory based on the extreme metastable states.
The lifetimes of the many-body metastable states would diverge with $N$, plausibly $\propto e^N$, and for large enough $N$, exceed the time accessible in numerical or experimental realizations.
We conjecture that an initial state with a finite correlation length
will lead, in the time window $1/\Gamma \ll t \ll 1/|\lambda_E|$, to one of the two mono-modal states $\rho_i$,
and {\em not} to an arbitrary combination of the two. A heuristic argument is given in \apporsm{App:MonomodalStates}. A product state is a natural reproducible initial state in an experiment, allowing to explore the metastability.
As a parameter is swept back and forth across the bistability region in an experimental setup, observables will show hysteretis loops -- unless the sweep is unrealistically slow ($\propto e^{-N}$).

{\it Experimental feasibility.} 
In addition to possible
realizations
with circuit-QED arrays \cite{PhysRevA.93.023821}, driven-dissipative spin
models
can be realized in current experiments with a few tens to a few hundreds of trapped ions. Ising and XY interactions can be implemented by laser beams inducing spin-motion coupling along one or two orthogonal directions \cite{PhysRevLett.92.207901,friedenauer2008simulating,Schneider_2012}, with an additional laser
for the on-site Hamiltonian. As recently demonstrated experimentally, the interaction can be varied from being almost independent of distance to
a dipolar power-law, and therefore short-range in 1D \cite{islam2013emergence,smith2016many} and 2D lattices \cite{bohnet2016quantum}. The interaction strength in these works is of order $J/\hbar\sim 10^4 s^{-1}$, one to two orders of magnitude larger than the qubit dephasing rates, and the rate of spin-flip processes in \eq{Eq:Dissipator} can be potentially controlled as well.

{\it To conclude,} 
studying lattices of driven-dissipative interacting spins using state-of-the-art 1D MPO simulations, for the parameters presented here and in further parameter regimes \cite{etatheory}, we have found no phase transition but a crossover between two regimes with different characteristics.
On the other hand, using a new approach that accounts for the leading-order lattice correlations and their feedback onto the mean magnetization, bistability appears to be possible in driven-dissipative quantum systems already in 2D.
Thus, the present exact and approximate calculations suggest that $D=2$ is a lower critical dimension for bistability in this problem.
This conclusion is consistent with works done in the context of Rydberg atoms on related models \cite{marcuzzi2014universal}, pointing toward a model-A dynamic universality class (whose lower critical dimension is known to be two) for the second order phase transition at the ending point of the bistability regime. This implies that in one dimension fluctuations destroy the critical point and with it the entire bistability region, in line with our results.

 The question of the existence of a lower critical dimension for bistability, bimodality and hysteresis and the accompanied dissipative phase transitions in this model can be directly addressed experimentally. If a definite answer is found, it would constitute the first demonstration of deciding a  question currently intractable classically, by a controlled quantum simulation. It could ascertain the status of the meanfield approximation in these systems, and shed light on the differences between equilibrium and nonequilibrium phase transitions.

\begin{acknowledgments}
We acknowledge the DRF of CEA for providing us with CPU time on the supercomputer COBALT of the CCRT. H.L. thanks Roni Geffen for fruitful discussions, and acknowledges support by IRS-IQUPS of Universit\'{e} Paris-Saclay and by LabEx PALM under grant number ANR-10-LABX-0039-PALM.

\end{acknowledgments}

%\clearpage
\appendix
\section{Bimodality}\label{App:Bimodality}

We start studying the model of Eqs.(1)-(3) of the main text on a finite \appredef{fully-connected}{FC} lattice with $N$ sites, i.e.~a graph where all sites are linked (with connectivity $\mathcal{Z}=N-1$). 
The FC version of the model is interesting because: (i) it has a unique steady state for any finite $N$, (ii) 
its meanfield (MF) solution gives a set of non-linear differential equations which can support bistability,
and (iii) the MF approximation becomes asymptotically exact in the limit $N\to\infty$. This is due to the fact that each site is coupled to the $x$ and $y$ components of the total magnetization of the $N-1$ other sites, and the fluctuations of this magnetization generically become small compared to its mean when $N\to\infty$.
As the thermodynamic limit $N\to\infty$ is approached, the bistability is thus expected to appear in some way
and the FC model can be viewed as a playground to investigate how a unique steady state at finite $N$ can be reconciled with bistability in the thermodynamic limit.

\begin{figure}
\includegraphics[width=3.3in]{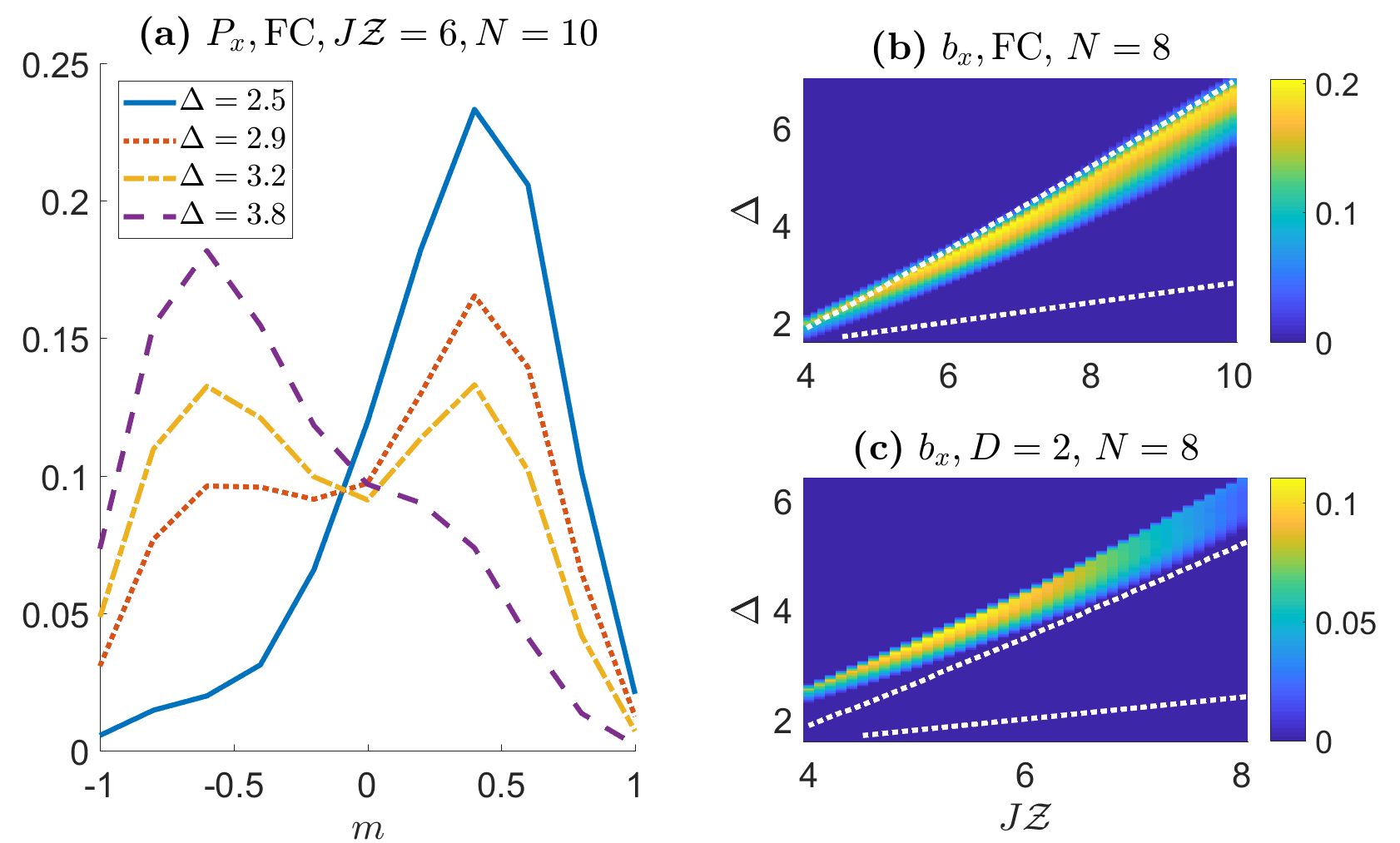}
\caption{(a) Probability distribution $P_x=P(M_x=m)$ of the $x$ magnetization per-site ($M_x=\frac{1}{N}\sum_R\sigma^x_R$), for a few values of the detuning $\Delta$, at fixed $J\mathcal{Z}$
and $\Omega=0.5$, in the steady state of the XY model with $N=10$ spins on a fully connected (FC) lattice. The transition from a single peak centered at $m>0$ to a single peak at $m<0$ as $\Delta$ is increased, is accompanied by an intermediate region of a bimodal distribution. (b) The bimodality index $b_x$ (see text) given by the color code, with $\Omega=0.5$ and $N=8$ for the FC lattice. Within the region bounded by the two white dotted lines, the meanfield limit manifests bistability of two different steady-state magnetizations. (c) $b_x$ for a 
two-dimensional ($2D$) parallelogram of $N=8$ sites and periodic boundary conditions (note the smaller $J\mathcal{Z}$ range). We find bimodality, although weaker than in the FC model, and with stronger finite-size effects.} \label{Fig:Bimodality}
\end{figure}

We thus compute exactly the unique stationary state density matrix $\rho_{ss}=\lim_{t\to\infty}\rho(t)$ on small systems. We focus on the probability distribution of the $x$ magnetization per site, i.e.~$P_x= P(M_x=m)$, where $M_x=\frac{1}{N}\sum_R\sigma_R^x$, that we plot in Fig.~\ref{Fig:Bimodality}(a) at fixed  $\Omega$ and $J\mathcal{Z}$ (the rescaling by $\mathcal{Z}$ allows to compare lattices of a different connectivity), and for different values of $\Delta$. Quite interestingly, such probability distribution evolves from a single peak centered around $m>0$ to a single peak centered around $m<0$, with an intermediate $\Delta$ range where $P_x$ has two separated peaks. To quantify the bimodality of the distribution
we define an index $b_x=2(P_{\max,2}-P_{\min})/(P_{\max,1}+P_{\max,2})$, where $P_{\max,2}\le P_{\max,1}$ are the two maxima of the distribution, and $P_{\min}$ is the minimum between the two. 
As can be seen in \fig{Fig:Bimodality}(b), the extent of the bimodality region in $\Delta$ and the maximum of $b_x$ increase with $J\mathcal{Z}$ at a fixed $\Omega$, following the MF bistability region.

While in the finite size limit we consider here the stationary state is still unique and the average value of the observable will either evolve smoothly with $\Delta$ or give rise to a step, depending on the scaling with $N$ of the peaks, the emergence of a bimodal probability distribution suggests a possible scenario for bistability to survive. Indeed for a FC lattice taking the thermodynamic limit first leads to a set of non-linear differential equations which can support multiple stable solutions. This should translate, in the probability distribution of the magnetization at finite size and finite time, and depending on the initial condition, in the emergence of a second peak on time scales exponentially in the system size such that, in the thermodynamic limit the switching from one solution to the 
other becomes exponentially suppressed.

\begin{figure}
{\includegraphics[clip,width=0.47\textwidth]{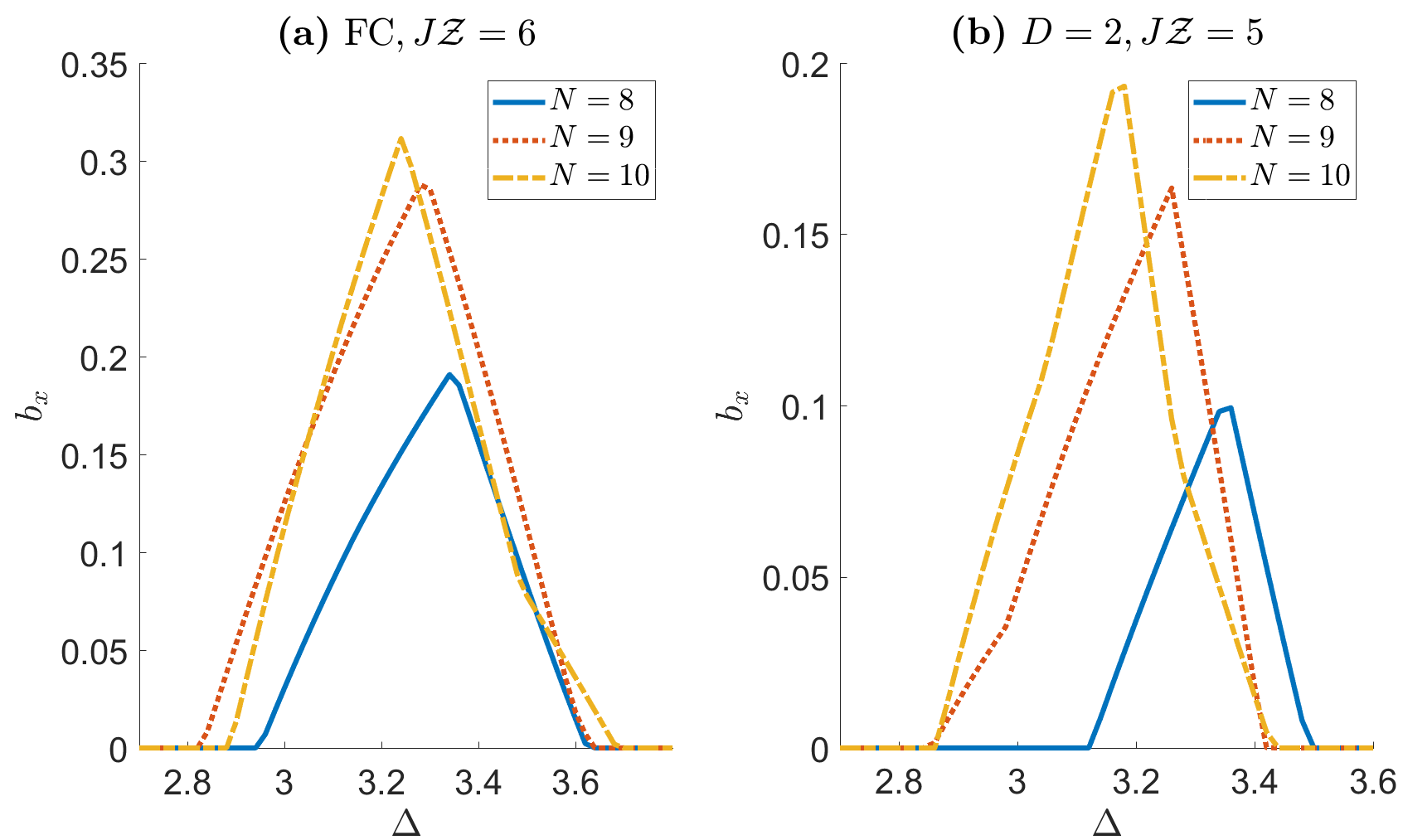}}
\caption{The bimodality index $b_x$ (see text) as a function of $\Delta$ for (a) $J\mathcal{Z}=6$ in a fully-connected (FC) lattice, and (b) $J\mathcal{Z}=5$ in a 2D lattice, from exact numerical solutions of the master equation of the driven-dissipative XY model, for small lattices, with three values of $N=8,9,10$. } \label{Fig:Biscaling}
\end{figure}

Repeating this analysis for a 2D square lattice with periodic boundary conditions (BC) we find again a bimodal parameter region, shifted towards higher $\Delta$, with and the maximum $b_x$ smaller than in the FC model [\fig{Fig:Bimodality}(c)]. With an increasing interaction strength $J$, the magnitude and extent of correlated fluctuations in the lattice grow, explaining the gradual decrease of the bimodal parameter region for a small lattice.
As further consistency checks, we verified that when increasing $N$ (up to 10) at specific parameter values, the bimodality range and maximum increase, and for a 1D chain we find (not shown) that $b_x$ remains strictly zero.

Figure \ref{Fig:Biscaling} shows the bimodality index $b_x$ calculated exactly for three increasing lattice sizes (in the \appredef{two-dimensional}{2D} case shown in panel (b), different parallelograms are constructed with periodic boundary conditions to incorporate $N$ sites), for the driven-dissipative XY model discussed in the main main text. Although sensitive to the finite system size, the width of the $\Delta$ ranges of bimodality increase with $N$ and the maximal $b_x$ value increases accordingly, which is consistent with bistability in the $N\to\infty$ limit.

\section{Bistability in a driven-dissipative Ising model in 2D}\label{App:Ising}

We have employed the \appredef{Meanfield with Quantum Fluctuations}{MFQF} method to the study of the driven dissipative Ising model, obtained after replacing the XY (flip-flop) interaction term in the Hamiltonian,
 \be H=\sum_{R}\left[\frac{\Delta}{2}\sigma_R^z+\Omega\sigma_R^x\right] -\sum_{\langle R,R'\rangle} \frac{J_z}{2} \sigma^z_R \sigma^z_{R'} ,\label{Eq:HIsing}\ee
while keeping the dissipator identical.
The equations of motion for $\vec{\mu}$ that are obtained from this model are identical to those in Eqs.~(4)-(6) of the main text with just the replacement $J\to -J_z$. The correlator equations (given in the following), are different.

\begin{figure}
{\includegraphics[clip,width=0.47\textwidth]{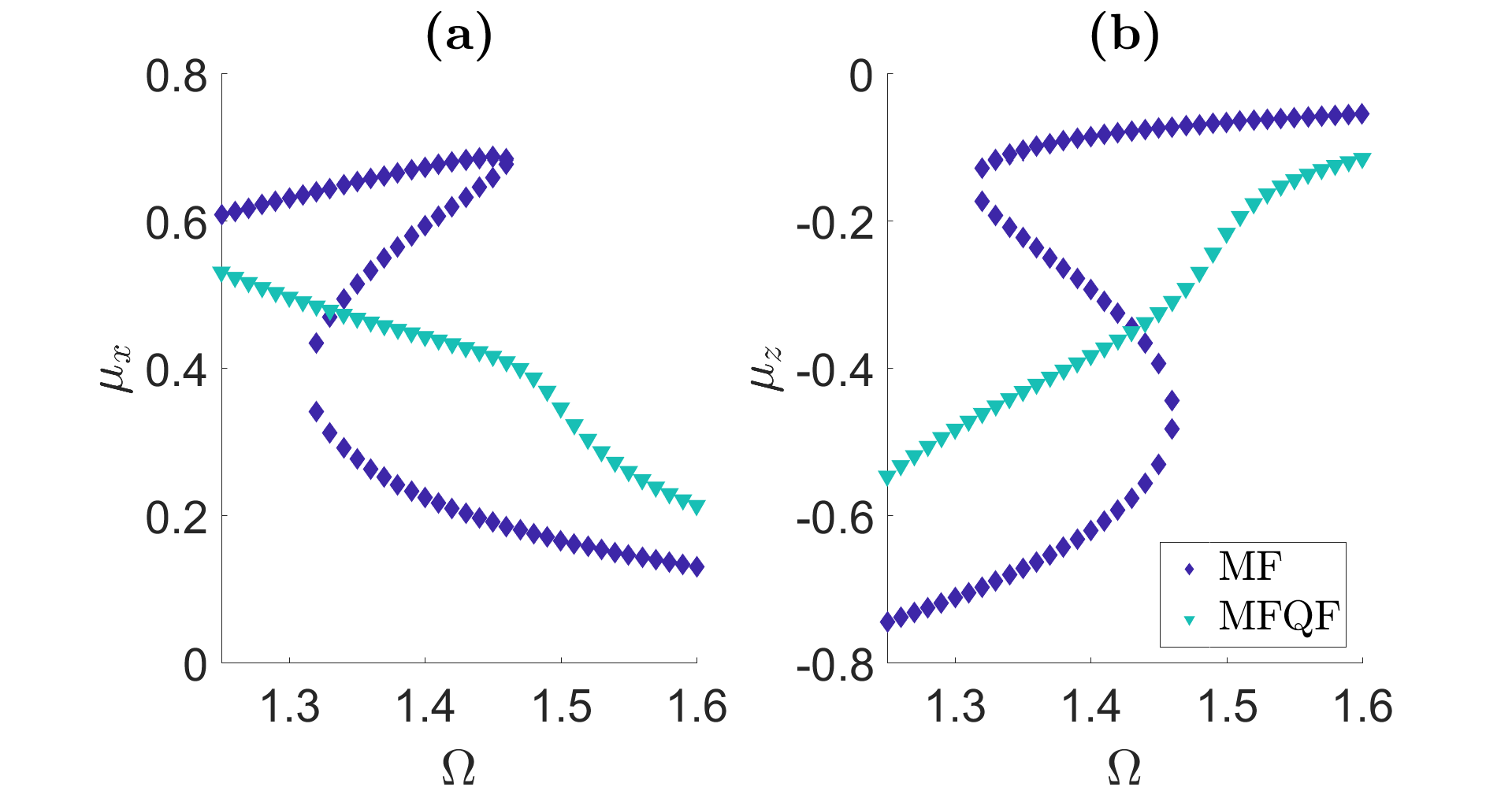}}
\caption{(a) Mean steady-state $x$ magnetization $ \mu_x^S$ as a function of $\Omega$ for $\Delta=0$ and $J_z\mathcal{Z}=-4$, on a 2D lattice for the Ising model. (b) The mean steady-state $z$ magnetization for the same parameters. The MF manifests bistability in the $\Omega$ range shown, while MFQF predicts a smooth crossover. We note that in the notation of Ref.~\cite{JinEtAlPRB18}, $\Omega=(h_x/\gamma)/2$, and $J_z\mathcal{Z}=-V/\gamma$ and the current figure corresponds to Fig.~1(a) of Ref.~\cite{JinEtAlPRB18}, in a somewhat larger $\Omega$ range, showing a rather sharp change around $\Omega\approx 1.5$. The simulation accounts for a lattice of $100^2$ sites, and the results have been verified to converge in $N$ and $t$.} \label{Fig:Ising2}
\end{figure}

Setting $\Delta=0$ we have studied the resulting MFQF steady state on a 2D lattice. A necessary condition for bistability in the meanfield limit \cite{etatheory}, is $|J_z\mathcal{Z}|>2$. For $J_z\mathcal{Z}=-4$, \fig{Fig:Ising2} shows a smooth crossover of the steady-state magnetization $\vec{\mu}^S$ as a function of $\Omega$ across the parameter region of meanfield bistability. The $z$ magnetization curve of \fig{Fig:Ising2}(b) can be compared with the results  plotted in Fig.~1(a) of \cite{JinEtAlPRB18}, which were obtained using a cluster MF approach for different cluster sizes. In the notation of Ref.~\cite{JinEtAlPRB18}, $J_z\mathcal{Z}=-V/\gamma$ and $\Omega=(h_x/\gamma)/2$, making the parameters in the plots identical (with only a somewhat larger range taken here for the abscissa). The MFQF curve that we obtain contains a further noticeable feature around $\Omega\approx 1.5$, with a relatively sharper change in magnetization.

\begin{figure}
{\includegraphics[clip,width=0.47\textwidth]{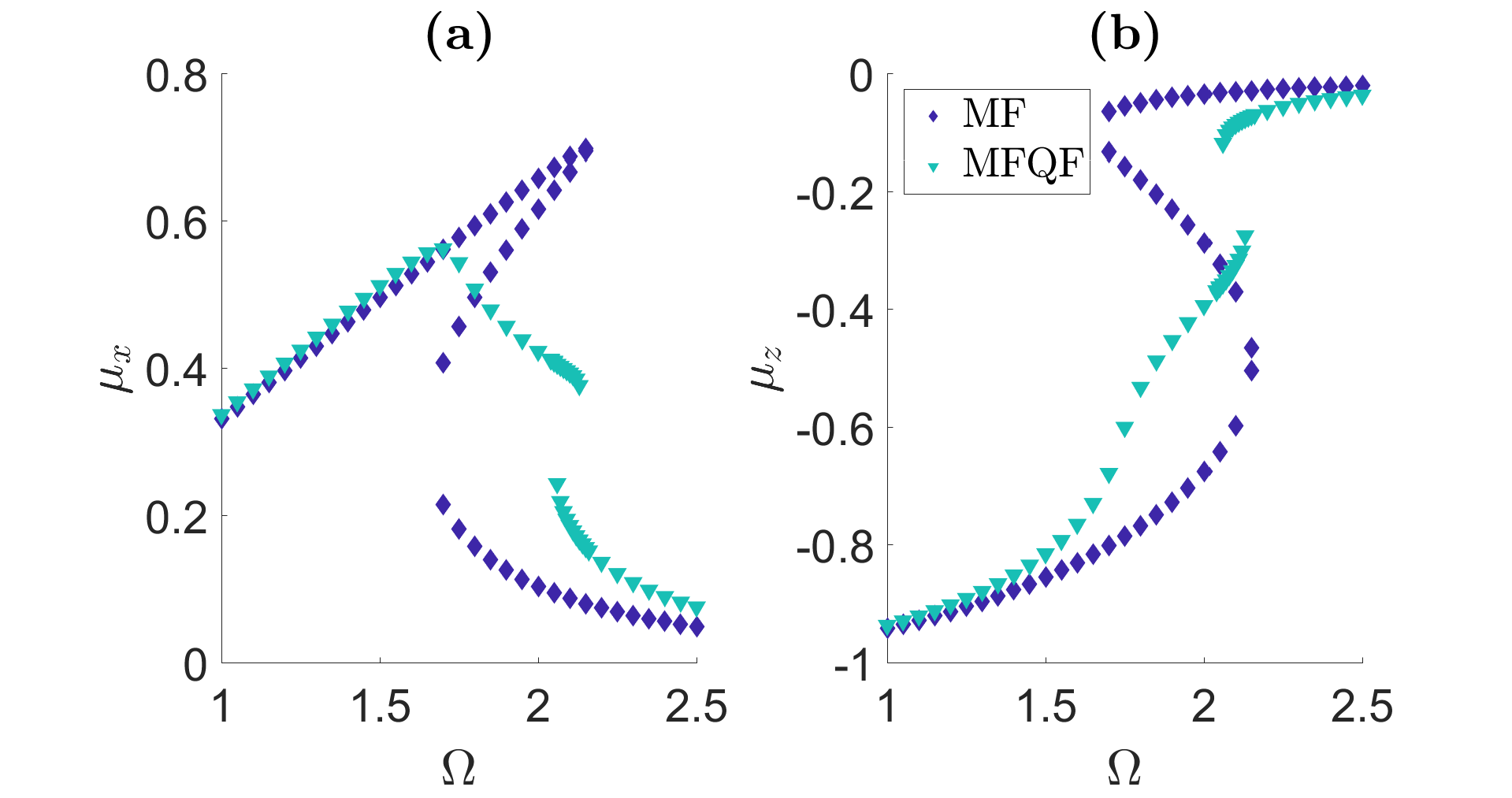}}
\caption{(a) Mean steady-state $x$ magnetization $ \mu_x^S$ as a function of $\Omega$ for $\Delta=0$ and $J_z\mathcal{Z}=-6$, on a 2D lattice for the Ising model. (b) The mean steady-state $z$ magnetization for the same parameters. The MF manifests bistability in the large $\Omega$ range shown, while MFQF predicts a shrinking of the bostability region and its occurrence at the edge of the $\Omega$ range. We note that in the notation of Ref.~\cite{JinEtAlPRB18}, $\Omega=(h_x/\gamma)/2$, and $J_z\mathcal{Z}=-V/\gamma$ and the current figure corresponds to Fig.~1(b) of Ref.~\cite{JinEtAlPRB18}. The simulation accounts for a lattice of $100^2$ sites, and the results have been verified to converge in $N$ and $t$. See \fig{Fig:CorrelationsIsing3} for the characteristics of corrleation functions around the bistable region.} \label{Fig:Ising3}
\end{figure}

\begin{figure}
{\includegraphics[clip,width=0.45\textwidth]{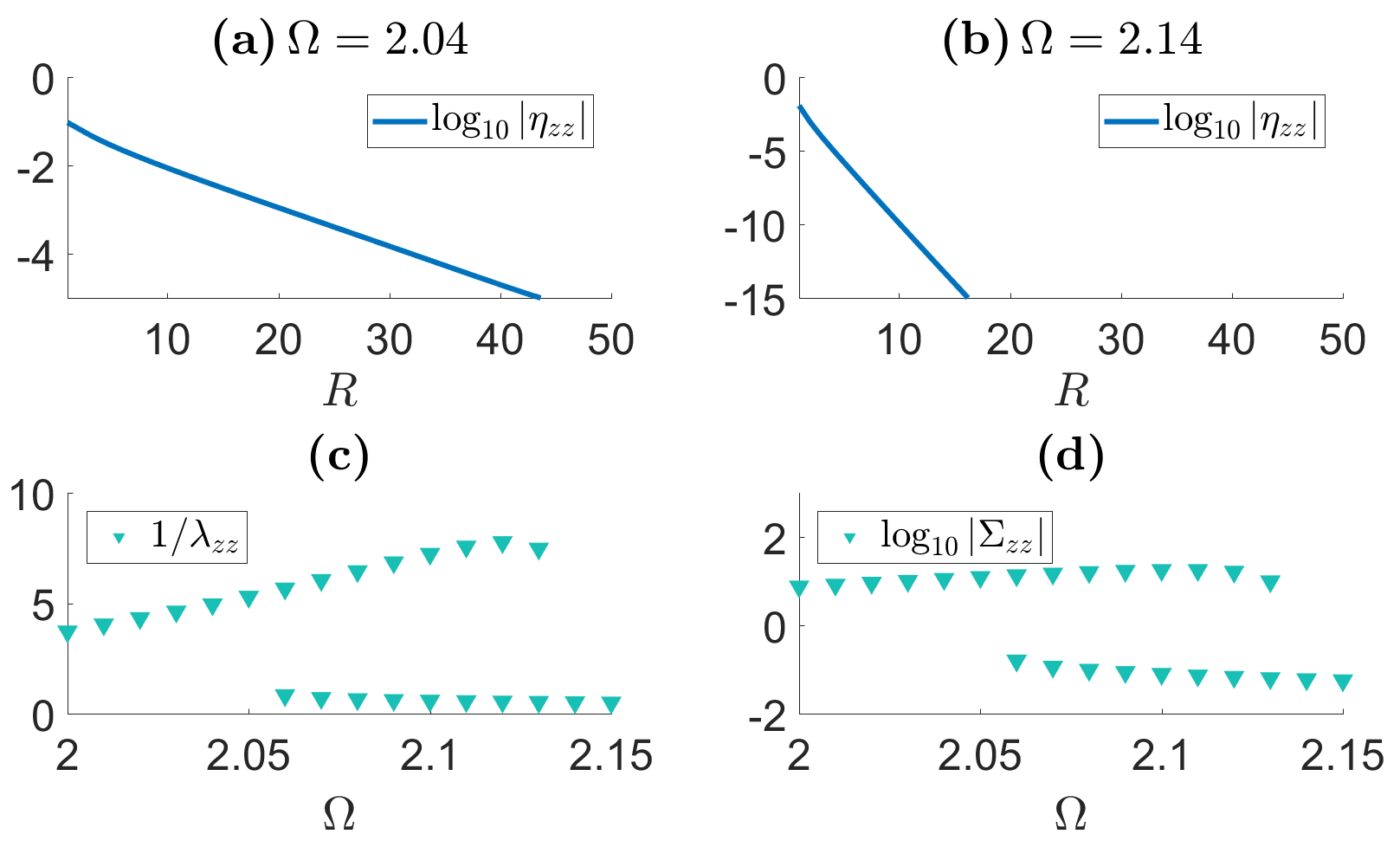}}
\caption{Characteristics of the correlation functions around the bistable region of \fig{Fig:Ising3}. (a)-(b) The correlation function $\eta_{zz}(R)$ as a function of the distance for two different $\Omega$ values (outside the bistability region). Note the different scales of the two plots. (c) The correlation length (along one spatial direction) as a function of $\Omega$, showing a difference of up to an order of magnitude in the bistable phases. (d) The total correlation $\Sigma_{zz}=\sum_R\eta_{zz}$, showing a difference of up to two orders of magnitude in the bistable phases.} \label{Fig:CorrelationsIsing3}
\end{figure}

In \fig{Fig:Ising3}, ${\mu}_x^S$ and ${\mu}_z^S$ are shown for $J_z\mathcal{Z}=-6$, in a larger region of $\Omega$ across the parameter region of meanfield bistability. The $z$ magnetization curve of \fig{Fig:Ising3}(b) can be compared with the curves plotted in Fig.~1(b) of \cite{JinEtAlPRB18}. In the region where the cluster MF of Ref.~\cite{JinEtAlPRB18} predicts a shrinking of the bistability (but not its complete disappearance up to the largest available cluster size), MFQF predicts a rather sharp but smooth crossover. However in MFQF approach bistability remains in a small region ($2.06\lesssim\Omega\lesssim 2.13$), at the right edge of MF bistability region. 
The structure of these the two co-existing stable phases is distinctly different in terms of the correlation functions $\eta_{ab}(R)$ defined in Eq.~(7) of the main text. Figure \ref{Fig:CorrelationsIsing3} presents the characteristics of $\eta_{zz}(R)$ as an example, showing the large differences in the correlation length, and -- as a result -- the total correlation, between the two bistable states.

\section{2D-MPO}\label{App:2D-MPO}

\begin{figure}
{\includegraphics[clip,width=0.47\textwidth]{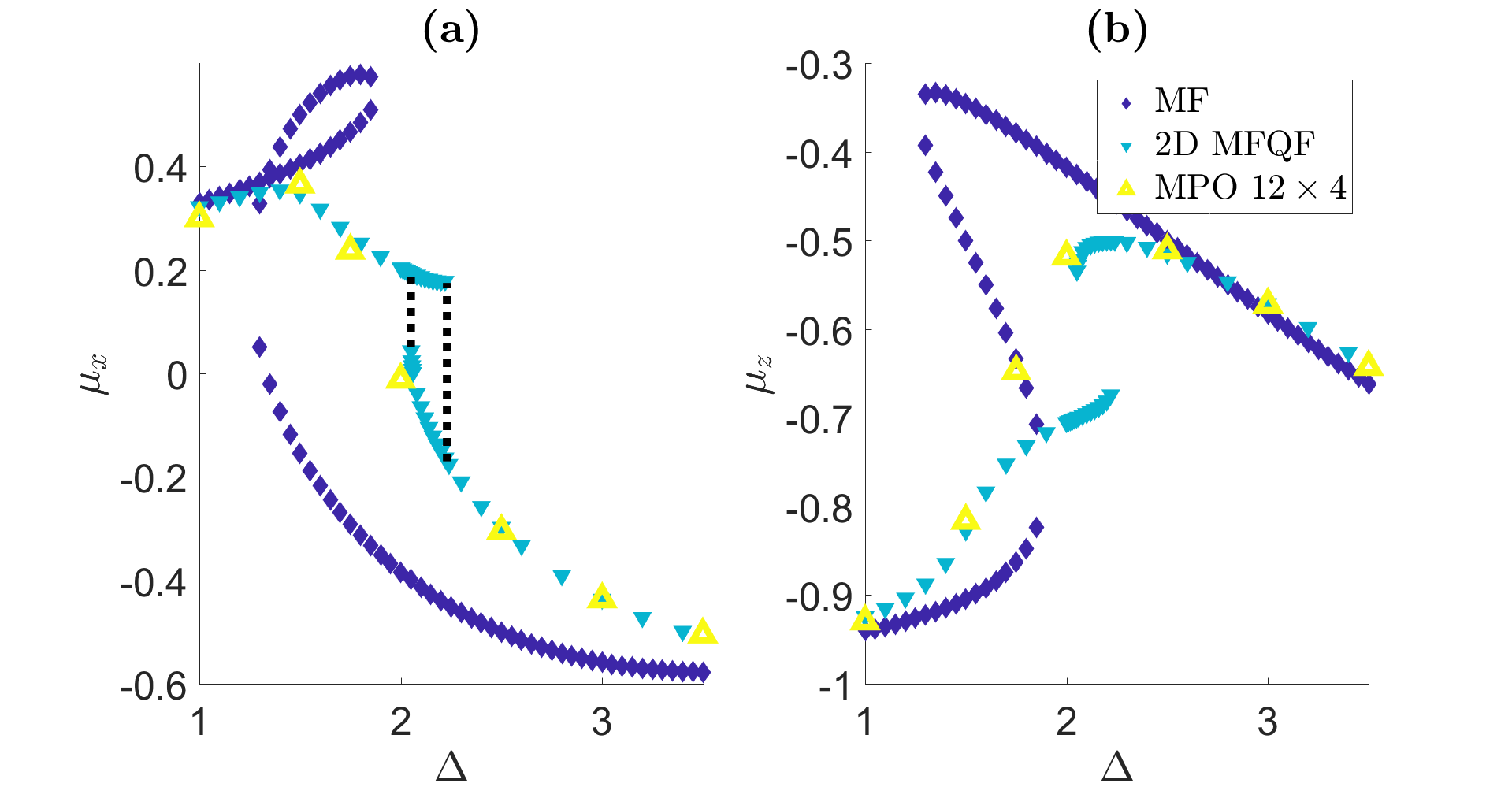}}
\caption{(a) Mean steady-state $x$ magnetization $\mu_x^S$ and (b) $z$ magnetization $\mu_z^S$, in the MFQF approximation in 2D, together with the MF limit and
2D-MPO results (for a finite-size, $12\times4$ cylinder). The parameters are as in Fig.~3 of the main text.} \label{Fig:Comparison2DMPO}
\end{figure}

The 2D MPO calculations presented in Fig.~3 of the main text and in \fig{Fig:Comparison2DMPO}, have been carried out on cylinders of length $L_x$ from 8 to 16, and fixed $L_y=4$.
At fixed precision the required bond dimension $\chi$ 
is expected to be constant  with $L_x$, but exponentially large in $L_y$.
The data is obtained with a bond dimension $\chi=400$
(pushed to $\chi=600$ at $\Delta=1.75$). As in the 1D-MPO result, the steady state is obtained by evolving $\rho$ in time from an initial
state where all the spins are pointing down. In practice a total time between $15$ and $25$ was used.
Since the path of interacting spins associated to the MPO artificially breaks the translation invariance of the lattice in the $y$ direction, it is
important to check that the bond dimension is large enough to restore the translation symmetry in the observables.
In our case the magnetization was found to be translation invariant up to relative errors of the order of $\mathcal{O}(10^{-4})$.

\section{Finite correlation length and mono-modal states}\label{App:MonomodalStates}

Consider a system where the Liouvilian spectrum has a steady state $\rho_{ss}$,
associated to the eigenvalue 0, and an eigenstate $\rho_E$ associated to an eigenvalue $\lambda_E$
which goes to zero when increasing the system size. We assume that all the other eigenstates are gapped
and can be ignored for times $t\gg 1/\Gamma$. As discussed in the main text, there exits two special linear combinations $\rho_{i=1,2}$
of $\rho_{ss}$ and $\rho_E$, which are the extreme states [56], and in which local observables have probability distributions with a single peak. On the other hand, nontrivial linear combinations
of $\rho_1$ and $\rho_2$ host bimodal distributions. A natural question is then: what initial conditions lead to $\rho_1$ and $\rho_2$, and what initial conditions instead lead to some convex combination of the two? We give below a simple heuristic argument suggesting that initial conditions where connected correlation functions have a finite correlation length, should, for a large enough system, lead to one of the two extreme states and {\em not} to arbitrary convex combinations of the two. Take an initial state $\rho(t=0)$ where all connected
correlation functions are short-ranged. Thanks to the assumption of a gap in the Liouvillian spectrum (above
$|\lambda_E|$),  the state $\rho(t)$ will reach the metastable manifold
  (spanned by $\rho_{ss}$ and $\rho_E$) rather quickly, after a time scale of order
$\mathcal{O}(N^0)$, {\it i.e.~}a time that is not growing with the system size. Because lattice models
with short-range interactions typically have a maximum (Lieb-Robinson)
velocity, beyond which the information cannot propagate, it
should not be possible for the system to develop true long-range
correlations (spanning the whole system) in a finite amount of time. And if correlations remain negligible at distances comparable with the system size, then this implies that the probability distributions
of local observable cannot be bi-modal. So, the Lindblad dynamics of a large system initialized from a state with short-range correlations
cannot lead to a bimodal state in the time range $1/\Gamma \ll t \ll N^{1/D}$ ($N^{1/D}$ is proportional to the linear size of the system). We thus conclude that in this time window the state that is reached is close to one of the two extreme states, since any other convex combination would be bimodal. 

\section{Meanfield with Quantum Fluctuations}\label{App:eta}

To derive the equations of the MFQF approach, we define a two-point correlation function (correlator),
\be \vartheta_{ab}(R,R',t)\equiv \left\langle \sigma_{R}^a\sigma_{R'}^b\right\rangle,\qquad R\neq R',\label{Eq:vartheta}\ee
which is a function of the difference $R-R'$ alone, symmetric in $a,b$ (because $\sigma_{R'}^a$ and $\sigma_R^b$ commute). Using \eq{Eq:vartheta}, the connected two-point correlator is defined as in Eq.~(7) of the main text, for $R\neq R'$, by
\be \eta_{ab}(R,R',t)=\vartheta_{ab}(R,R') - \mu_a\mu_b,\label{Eq:etadef}\ee
The connected three-point correlator is defined for $R\neq R'\neq R''$ by
\be \zeta_{abc}(R,R',R'',t) \equiv \left\langle \left(\sigma_{R}^a -\mu_a\right)\left(\sigma_{R'}^b-\mu_b\right)\left(\sigma_{R''}^c-\mu_c\right)\right\rangle,\label{Eq:tilderhoR0}
\ee
which is again a function of the differences only.

The approximation of the following treatment is based on assuming that $\zeta$ (and higher order connected correlators) can be neglected in comparison to $\eta$. The e.o.m of $\vartheta_{ab}$, setting $R'=0$, is
\bem  \partial_t\vartheta_{ab}(R)=\\ \sum_d\Pi_{ad}\vartheta_{db}(R) + \sum_d\Pi_{bd}\vartheta_{ad}(R)+ f_{ab} (\mu,\vartheta)+ g_{ab} (\mu,\vartheta), \label{Eq:dotetar_ab}\end{multline}
where the local Hamiltonian terms are described using the matrix
\be  \Pi= \left(\begin{array}{ccc}
 0 & -\Delta & 0
  \\
  \Delta & 0 & -2{\Omega} 
  \\
0 & 2{\Omega} &  0
   \end{array}\right),\ee
while $f_{ab}(\mu,\vartheta)\propto J$ comes from the kinetic terms, and $g_{ab}(\mu,\vartheta)\propto \Gamma$ comes from the Lindbladian part, and both are given below. By using \eq{Eq:etadef} we get the e.o.m system for $\eta(R,t)$,
\be \partial_t \eta_{ab}(R,t)=\partial_t \vartheta_{ab}(R) - \partial_t \left[\mu_a\mu_b\right],\label{Eq:etaeom}\ee
which we solve numerically together with the coupled system for $\vec\mu(t)$.

For the Hamiltonian in Eq.~(2) of the main text where the kinetic term is
\be -\sum_{\langle R,R'\rangle} J\left(\sigma^+_R \sigma^-_{R'} +{\rm h.c.}\right) =-\sum_{\langle R,R'\rangle} \frac{J}{2}\left(\sigma^x_R \sigma^x_{R'} +\sigma^y_R \sigma^y_{R'}\right) ,\ee
we find that $f_{ab} (\mu,\vartheta)$ of \eq{Eq:dotetar_ab} is given by,
\begin{widetext} 
\be
f_{xx}(R)= 2J \left[2 \mu_x\mu_y\mu_z -  \mu_x \vartheta_{yz}(1) -\mu_y \vartheta_{xz}(R)\right]\left[\mathcal{Z}-\delta_{\|R\|,1}\right] - 2J\sum_{\substack{
R'\neq 0 \\ \| R'- R\|=1}}\mu_z \vartheta_{xy}(R'),
\ee
\be
f_{yy}(R)= -2J \left[2 \mu_x\mu_y\mu_z -  \mu_y \vartheta_{xz}(1) -\mu_x \vartheta_{yz}(R)\right]\left[\mathcal{Z}-\delta_{\|R\|,1}\right] + 2J\sum_{\substack{
R'\neq 0 \\ \| R'- R\|=1}}\mu_z \vartheta_{xy}(R') ,
\ee
\be
f_{zz}(R)=  -2J \left[ \mu_x \vartheta_{yz}(R) -\mu_y\vartheta_{xz}(R) \right]\left[\mathcal{Z}-\delta_{\|R\|,1}\right] - 2J\sum_{\substack{
R'\neq 0\\\| R'- R\|=1}}\left[ \mu_y \vartheta_{xz}(R')- \mu_x \vartheta_{yz}(R')\right] .
\ee
\bem
f_{xy}(R)= J  \left[ 2\mu_y^2 \mu_z-2 \mu_x^2\mu_z -\mu_y \vartheta_{yz}(1) -\mu_y \vartheta_{yz}(R) + \mu_x \vartheta_{xz}(1) +\mu_x \vartheta_{xz}(R)\right]\left[\mathcal{Z}-\delta_{\|R\|,1}\right] \\ - J\sum_{\substack{
R'\neq 0\\\| R'- R\|=1}}\left[ \mu_z \vartheta_{yy}(R')  -\mu_z \vartheta_{xx}(R')\right],
\end{multline}
\pagebreak \clearpage
\bem
f_{xz}(R)= -J\mu_y\delta_{\|R\|,1} + J \left[ 2 \mu_z^2 \mu_y -\mu_z \vartheta_{yz}(1) -\mu_y \vartheta_{zz}(R) -\mu_x \vartheta_{xy}(R) +\mu_y \vartheta_{xx}(R) \right]\left[\mathcal{Z}-\delta_{\|R\|,1}\right] \\ - J\sum_{\substack{
R'\neq 0\\\| R'- R\|=1}}\left[ \mu_z \vartheta_{yz}(R') + \mu_y \vartheta_{xx}(R') - \mu_x\vartheta_{xy}(R') \right],
\end{multline}
\bem
f_{yz}(R)= J\mu_x \delta_{\|R\|,1} + J \left[ -2 \mu_z^2 \mu_x +\mu_z \vartheta_{xz}(1) +\mu_x \vartheta_{zz}(R) -\mu_x \vartheta_{yy}(R)+\mu_y \vartheta_{xy}(R)\right] \left[\mathcal{Z}-\delta_{\|R\|,1}\right] \\ + J\sum_{\substack{
R'\neq 0\\\| R'- R\|=1}}\left[ \mu_z \vartheta_{xz}(R') -\mu_y \vartheta_{xy}(R') + \mu_x\vartheta_{yy}(R') \right].
\end{multline}
The components of $g (\mu,\vartheta)$ in \eq{Eq:dotetar_ab} are given by
\be g_{aa} = -\Gamma\vartheta_{aa},\quad g_{xy}=-\Gamma\vartheta_{xy},\quad g_{xz}=-\Gamma\left[2\vartheta_{xz}+\frac{3}{2}\mu_x \right],\quad g_{yz}=-\Gamma\left[ 2\vartheta_{yz}+\frac{3}{2}\mu_y \right].\ee
For the Ising model of \eq{Eq:HIsing} where the kinetic term is
 \be -\sum_{\langle R,R'\rangle} \frac{J_z}{2} \sigma^z_R \sigma^z_{R'} ,\ee
 we get instead of the above $f_{ab} (\mu,\vartheta)$, the following expressions;
\be
f_{xx}(R)= -2J_z \left[2 \mu_x\mu_y\mu_z -  \mu_x \vartheta_{yz}(1) -\mu_z \vartheta_{xy}(R)\right]\left[\mathcal{Z}-\delta_{\|R\|,1}\right] + 2J_z \sum_{\substack{
R'\neq 0 \\ \| R'- R\|=1}}\mu_y \vartheta_{xz}(R'),
\ee
\be
f_{yy}(R)= 2J_z \left[2 \mu_x\mu_y\mu_z -  \mu_y \vartheta_{xz}(1) -\mu_z \vartheta_{xy}(R)\right]\left[\mathcal{Z}-\delta_{\|R\|,1}\right] - 2J_z \sum_{\substack{
R'\neq 0 \\ \| R'- R\|=1}}\mu_x \vartheta_{yz}(R') ,
\ee
\be
f_{zz}(R)=  0 .
\ee
\bem
f_{xy}(R)= J_z  \left[ 2\mu_x^2 \mu_z-2 \mu_y^2\mu_z -\mu_x \vartheta_{xz}(1) -\mu_z \vartheta_{xx}(R) + \mu_y \vartheta_{yz}(1) +\mu_z \vartheta_{yy}(R)\right]\left[\mathcal{Z}-\delta_{\|R\|,1}\right] \\ + J_z \sum_{\substack{
R'\neq 0\\\| R'- R\|=1}}\left[ \mu_y \vartheta_{yz}(R')  -\mu_x \vartheta_{xz}(R')\right],
\end{multline}
\be
f_{xz}(R)= J_z\mu_y\delta_{\|R\|,1} - J_z \left[ 2 \mu_z^2 \mu_y -\mu_z \vartheta_{yz}(1) -\mu_z \vartheta_{yz}(R) \right]\left[\mathcal{Z}-\delta_{\|R\|,1}\right] \\ + J_z\sum_{\substack{
R'\neq 0\\\| R'- R\|=1}}\mu_y \vartheta_{zz}(R'),
\ee
\be
f_{yz}(R)= -J_z\mu_x \delta_{\|R\|,1} + J_z \left[ 2 \mu_z^2 \mu_x -\mu_z \vartheta_{xz}(1) -\mu_z \vartheta_{xz}(R)\right] \left[\mathcal{Z}-\delta_{\|R\|,1}\right] \\ - J_z\sum_{\substack{
R'\neq 0\\\| R'- R\|=1}}\mu_x \vartheta_{zz}(R') .
\ee

\pagebreak \clearpage

\end{widetext}

\bibliography{nonequilibrium}

\end{document}